\documentclass[12pt]{article}
\usepackage{amsmath, amssymb}
\usepackage[dvips]{epsfig}

\advance \textwidth by -18 mm

\def\##1{\underline{#1}}
\def\=#1{\underline{\underline{#1}}}

\def\+#1{\underline{\bf #1}}
\def\*#1{\underline{\underline{\bf #1}}}

\def\.{\mbox{ \tiny{$^\bullet$} }}

\def\eps{\epsilon}
\def\epso{\epsilon_{\scriptscriptstyle 0}}
\def\muo{\mu_{\scriptscriptstyle 0}}

\def\ko{k_{\scriptscriptstyle 0}}
\def\lambdao{\lambda_{\scriptscriptstyle 0}}
\def\etao{\eta_{\scriptscriptstyle 0}}

\def\epsa{\eps_a}
\def\mua{\mu_a}
\def\epsb{\eps_b}
\def\mub{\mu_b}

\def\ux{{\bf u}_x}
\def\uy{{\bf u}_y}
\def\uz{{\bf u}_z}

\def\c#1{\cite{#1}}

\def\le{\left(}
\def\ri{\right)}
\def\les{\left[}
\def\ris{\right]}
\def\lec{\left\{}
\def\ric{\right\}}
\begin{document}

\noindent {\bf Negative Index of Refraction and Distributed Bragg Reflectors}

\bigskip
\noindent Jaline Gerardin$^{1}$ and Akhlesh Lakhtakia$^{1}$\\

\bigskip

\noindent $^{1}${ CATMAS --- Computational and Theoretical Materials Sciences Group\\
     Department of Engineering Science and Mechanics\\
     Pennsylvania State University, University Park, PA
     16802--6812}

\bigskip

\noindent {\bf ABSTRACT:} The Bragg regime shifts when
conventional materials in a
multilayer distributed Bragg reflector (DBR)
are replaced by artificial
materials with the so--called negative index of refraction. 
This provides an avenue for characterizing the latter class
of materials.
\\


\newpage

\section{INTRODUCTION}

Artificial
materials with the so--called {\em negative index of refraction\/}
at $\sim$ 10~GHz frequency 
have garnered much recent interest  \c{SK}--\c{Jian}. 
These isotropic materials, with supposedly 
negative real permittivity and negative real permeability, were the subject of
a mid--1960s paper of Veselago \c{Ves}, who 
predicted certain unusual electromagnetic properties and
coined the unclear term
{\em left--handed materials\/} for them. Handed these materials
are not \c{Bel},
 issue can be taken on the isotropy of their first samples \c{SSS},
and  strictly non--dissipative no materials can be 
\c{WL}.
Yet the only available experimental result \c{SSS}
conclusively shows that these materials are different from
their conventional counterparts (i.e., those with positive real permittivity
and positive real permeability). As better realizations
are probably in the wings, in colloquial terms, the business
of macroscopic electromagnetics can no longer be as usual.

The manufacturing process of the subject artificial materials 
delivers samples in the form
of slabs.
Even a cursory perusal of key papers \c{SK}--\c{SSS}
reveals that 
characteristic observable properties of these
materials  supposedly are manifested most clearly when 
plane waves are obliquely incident on  planar interfaces
with  their conventional counterparts. Indeed, the only
experimental confirmation of their unusual characteristics
exploited oblique incidence on a planar interface with
air \c{SSS}. In contrast, we present here a configuration
wherein the incidence is normal and yet the subject
artificial materials can be easily distinguished from
their conventional counterparts as follows.

\section{ANALYSIS}
Distributed Bragg reflectors (DBRs)
are commonly used planar devices
 in optics \c{Mac}--\c{GP}. A multilayer DBR
is a stack of layers of two alternating materials
with different  indexes of refraction as well as low absorption, exhibiting
very high reflectance in the so--called Bragg regime.
If $P$ denotes the thickness of a unit cell
comprising two adjacent layers, and $\bar{n}$
is some {\em effective\/}  index of refraction, then the Bragg regime
is located about the (free--space) wavelength \c{deG}--\c{Oth}
\begin{equation}
\label{Br}
\lambdao^{Br} = 2P\bar{n}\,.
\end{equation}
(Bragg regimes are also possible around integer
submultiples of this $\lambdao^{Br}$, but it must be borne in mind
that  indexes of refraction are 
wavelength--dependent, in general.)
The parameter $\bar n$ depends
on the indexes of refraction and the volumetric proportion
of the two constituent materials.

Suppose
the region $0\leq z\leq L$ is occupied by a multilayer DBR, as
shown in Figure 1. The multilayer DBR  comprises 
$N$ unit cells, each made of two 
layers labeled $a$ and $b$, with relative permittivities $\eps_{a,b}$
and relative permeabilities $\mu_{a,b}$. With $P = L/N$ as the
thickness of the
unit cell, the thickness of layer $a$ is equal to $qP$, 
$0\leq q\leq 1$.
A plane wave is normally incident on the DBR from the
vacuous half--space
$z \leq 0$, with $\lambdao$ denoting its
wavelength. Therefore, a reflected plane wave also exists
in the half--space $z\leq 0$, and a transmitted plane wave
in the vacuous half--space $z\geq L$. The corresponding electric field
phasors are given by
\begin{equation}
{\bf E}(z,\lambdao) = \ux\,
\lec
\begin{array}{ll}
a \exp(i\ko z) + r\exp(-i\ko z)\,,&\qquad z \leq 0
\\
t \exp \les i\ko (z-L)\ris\,,&\qquad z \geq L
\end{array}\right.,
\end{equation}
where $\ko=2\pi/\lambdao$ is the free--space
wavenumber; $a$, $r$ and $t$ are the amplitudes
of the incident, reflected and transmitted plane waves, respectively;
while $(\ux,\uy,\uz)$ is the triad of cartesian unit vectors.
An $\exp(-i\omega t)$ time--dependence is implicit, where $\omega
=\ko /(\epso\muo)^{1/2}$
is the angular frequency, while $\epso$
and $\muo$ are the permittivity and the permeability
of free space, respectively.

The amplitudes $r$ and $t$ must be determined in terms of $a$. This
is best done by setting
up the 2$\times$2 matrix equation
\c{Lopt}
\begin{eqnarray}
\nonumber
t \, \les\begin{array}{c} 1 \\ \etao^{-1} \end{array}\ris &=&
\le
\exp  \lec i\omega (1-q)P \les\begin{array}{cc} 0 & \muo\mu_b \\
\epso\eps_b & 0 \end{array}\ris\ric \right.
\\
& &\qquad
\left.
\exp  \lec i\omega qP \les\begin{array}{cc} 0 & \muo\mu_a \\
\epso\eps_a & 0 \end{array}\ris\ric
\ri^N
\les\begin{array}{c} (a+r) \\\etao^{-1} (a-r) \end{array}\ris\,,
\end{eqnarray}
where $\etao =  (\muo/\epso)^{1/2}$ is the intrinsic
impedance of free space. This equation has to be
numerically solved, which we did. The principle of conservation of energy
requires that $\vert r\vert^2 + \vert t\vert^2 \leq \vert a\vert^2$,
with the equality holding
only if the both constitutent materials
in the DBR are non--dissipative at the particular value of $\lambdao$.
Our algorithm satisfied the conservation principle. 

\section{NUMERICAL RESULTS}
Figure 2 shows the computed reflectance $\vert r/a\vert^2$ as a
function of $\lambdao/P$ for three
values of $q$ when $\epsa=4(1+i0.001)$, $\mua=1.02(1+i0.001)$,
$\epsb=1$, $\mub=1$ and $N=20$. The Bragg regime is clearly observable
via the  rectangular feature with an almost flat top and two vertical sides
in all three plots. The 
full--width--at--half--maximum bandwidth $\Delta\lambdao$ of the Bragg regime
is typically $< 1.25\,P$ for all $q$.
Predictably, the Bragg feature vanishes for $q=0$
and $q=1$. As $q$ increases, so does $\lambdao^{Br}/P$ (at the
center of the Bragg feature in each plot); which 
amounts to an increase in $\bar n$, as shown in Table 1.

Ideal Bragg features  do not emerge for all values of
$q \in \les 0.2,\, 0.5\ris$, 
when $\epsa=4(-1+i0.001)$, $\mua=1.02(-1+i0.001)$,
$\epsb=1$, $\mub=1$ and $N=20$. Thus, the Bragg feature
for $q=0.5$ is not flat--topped in Figure 3,
although it is well--developed 
for $q=0.6$ and $q=0.7$. Calculated values
of $\bar n$ as functions
of $q$ are  shown in
Table 1. 

We note from the presented and related results
 that $\bar n > 1$ for all values of $q \in (0,\,1)$ when
${\rm Re}[\epsa,\mua] >0$. In contrast,
 $0 <\bar n < 1$ for $q \stackrel{<}{\sim} 0.67$
when ${\rm Re}[\epsa,\mua] <0$. The reduction of $\bar n$ below
the unit index of  refraction of material $b$ 
could suggest that
the real part of the index of refraction is negative
for the subject artificial materials, but that suggestion
does not appear to be supported by the values of $\bar n > 1$ for
 $q \stackrel{>}{\sim} 0.67$
when ${\rm Re}[\epsa,\mua] <0$.  Anyhow,
in conjunction
with Figure 3,  Table 1 confirms
that wave--material interaction in the subject artificial
materials is intrinsically different from that in their
conventional counterparts.

Our results also show that  the Bragg regime
would shift to shorter wavelengths, if a conventional dielectric/magnetic
constituent of a multilayer DBR were to be replaced by
its analog of the subject variety. 
Consequently, measurements of $\bar n$ would
illuminate the issue of the {\em negative\/}  index of
refraction, and could  also
help in the characterization of the subject
artificial materials. At the same time,
multilayer DBRs made with the subject
artificial materials could be useful in 
wavelength regimes that are inaccessible with DBRs
made with only conventional materials.

\newpage
\begin{center}

{\bf Table 1} Values of ${\bar n} = \lambdao^{Br}/2P$
computed for different values
of $q$; $N=20$,
$\epsb=1$ and $\mub=1$.
  
\begin{tabular}{|c||c|c|}
\hline
$q$ & $\epsa=4(1+i0.001)$  & $\epsa=4(-1+i0.001)$ \\
 &  $\mua=1.02(1+i0.001)$ & $\mua=1.02(-1+i0.001)$\\

\hline
0.1 & 1.14 & 0.69\\
0.2 & 1.26 & 0.37$^\dag$ \\
0.3 & 1.38 & 0.09$^\dag$ \\
0.4 & 1.47 & 0.23$^\dag$\\
0.5 & 1.58 & 0.56$^\dag$\\
0.6 & 1.68 & 0.82 \\
0.7 & 1.78 & 1.09\\
0.8 & 1.87 & 1.37\\
0.9 & 1.96$^\dag$ & 1.68$^\dag$ \\
\hline

\end{tabular}
\end{center}
\smallskip
\noindent $^\dag$  Bragg feature has a curved top~---~exemplified
in the top plot of Figure 3~---~which begins to flatten as $N$
increases. There exists also a certain degree of arbitrariness in
the identification of the Bragg feature for some values of $q$
between $\sim 0.25$ and ${\sim}0.42$ when 
${\rm Re}[\epsa] =-4$ and ${\rm Re}[\mua] =-1.02$.


\newpage

\begin{figure}[!ht]
\centering \psfull \epsfig{file=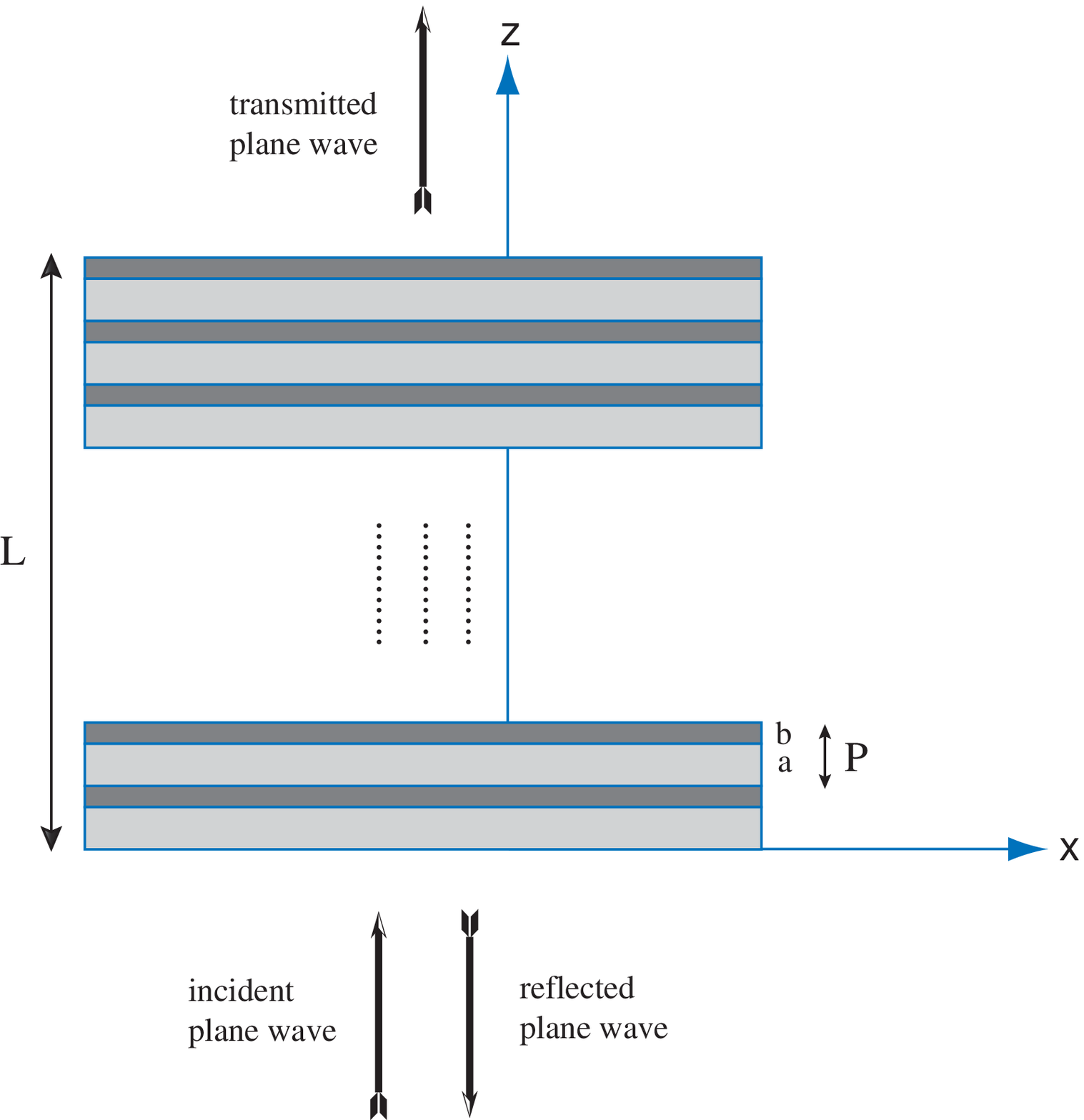,width=3in}
\end{figure}
\bigskip\bigskip
\noindent Figure 1:
Schematic of the boundary value problem. The distributed
Bragg reflector comprises $N$ unit cells, each of thickness
$P$ and containing one layer each of two different materials
labeled $a$ and $b$.

\newpage

\begin{figure}[!ht]
\centering \psfull \epsfig{file=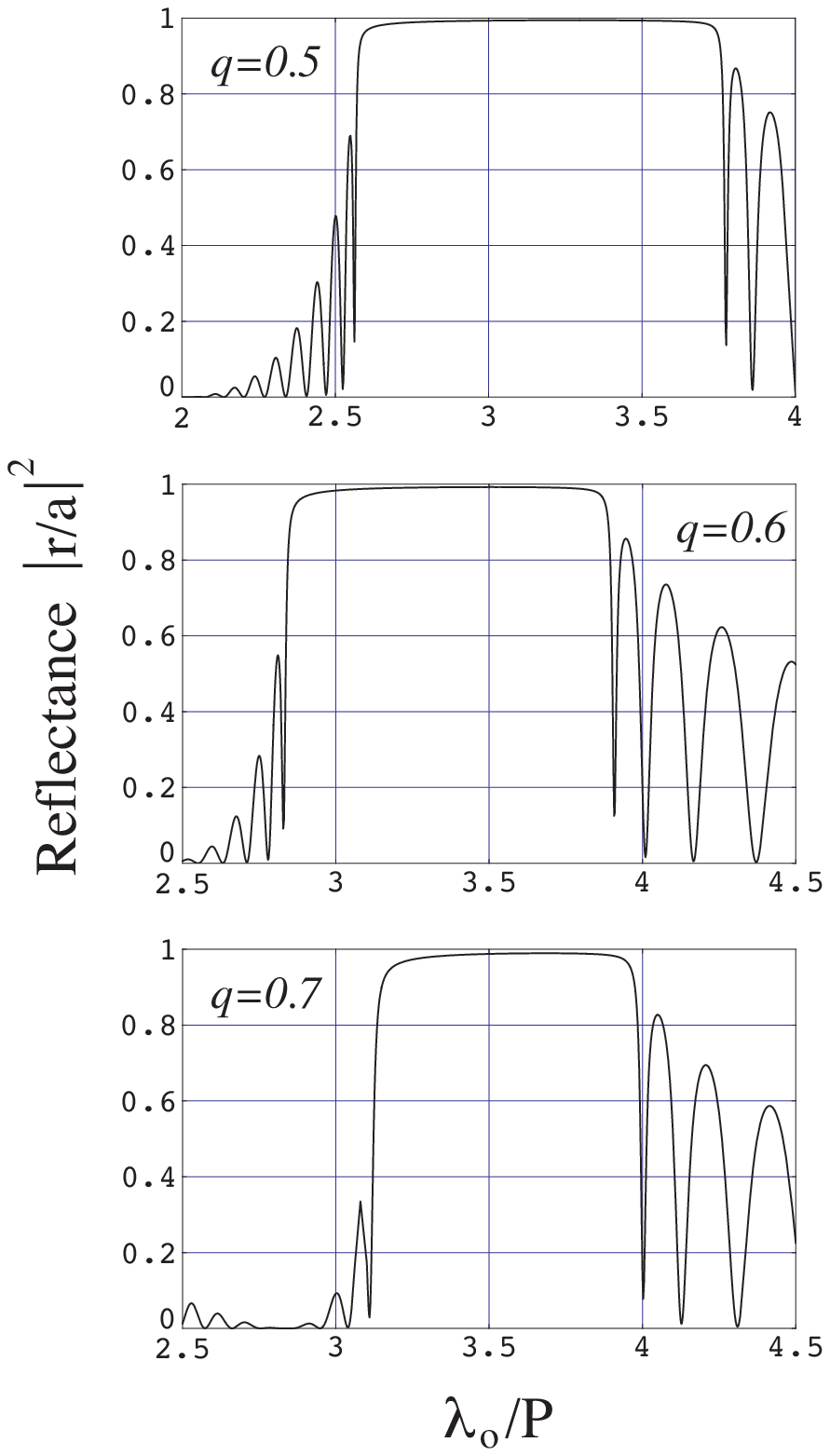,width=3in}
\end{figure}
\bigskip\bigskip
\noindent Figure 2:
Reflectance $\vert r/a\vert^2$ of a DBR as a function of
$\lambdao/P$ for different values
of $q$; $N=20$, $\epsa=4(1+i0.001)$, $\mua=1.02(1+i0.001)$,
$\epsb=1$ and $\mub=1$.

\newpage

\begin{figure}[!ht]
\centering \psfull \epsfig{file=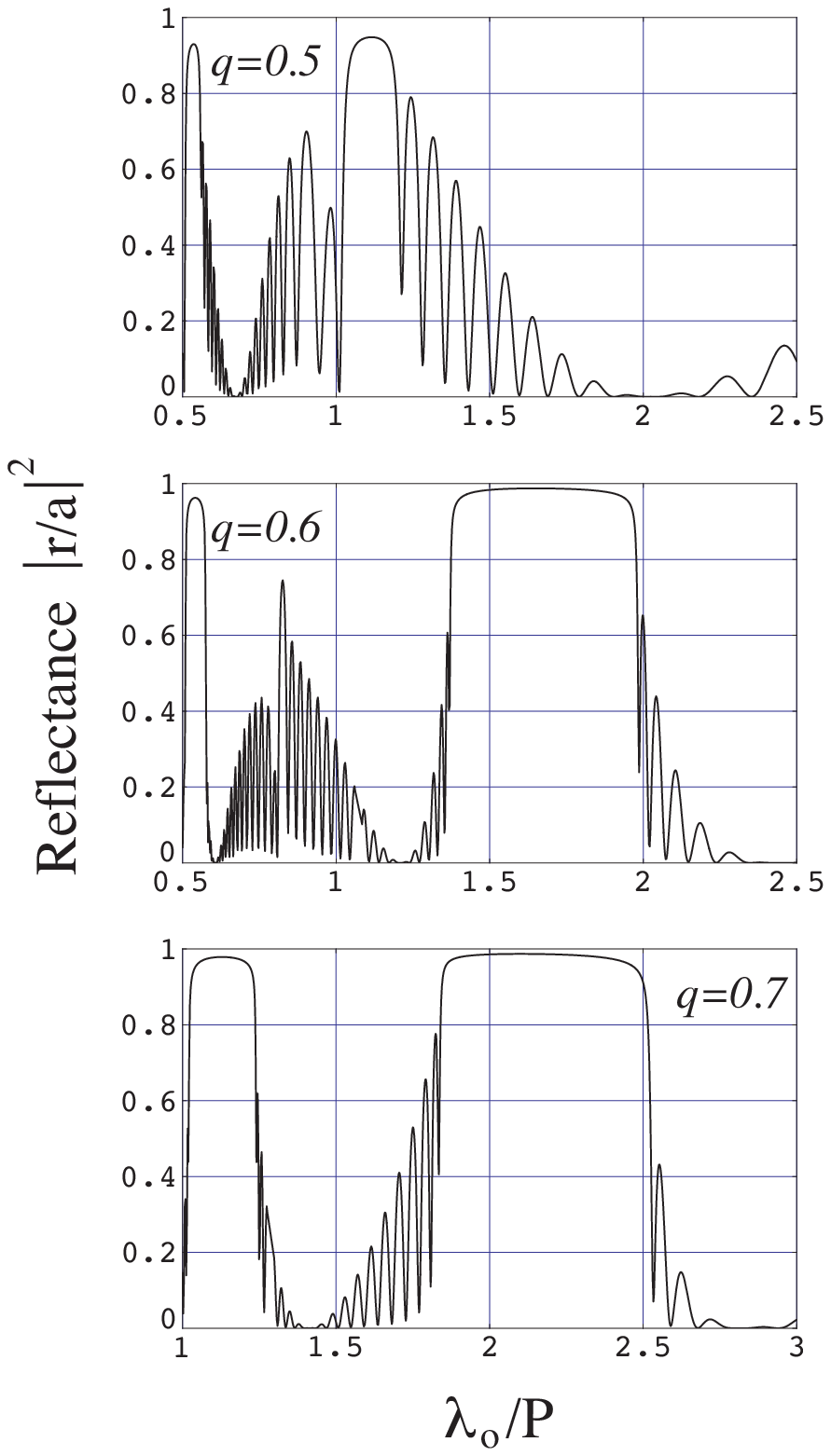,width=3in}
\end{figure}
\bigskip\bigskip
\noindent Figure 3:
Same as Figure 2, except $\epsa=4(-1+i0.001)$ and $\mua=1.02(-1+i0.001)$.

\end{document}